\documentclass[aps,final,notitlepage,oneside,twocolumn,nobibnotes,nofootinbib,
superscriptaddress,noshowpacs,centertags]{revtex4-1}

\usepackage[english]{babel}
\usepackage{graphicx}
\usepackage{latexsym}
\usepackage{amssymb}
\usepackage{amsmath}
\usepackage{float}

\usepackage{xcolor}

\begin{document}

\title{Possible Explanation of the Geograv Detector Signal during the Explosion of SN 1987A in Modified Gravity Models}
\author{Yu. N. Eroshenko}\thanks{e-mail: eroshenko@inr.ac.ru}
\affiliation{Institute for Nuclear Research, Russian Academy of Sciences, pr. 60-letiya Oktyabrya 7a, Moscow, 117312 Russia}
\author{E. O. Babichev}\thanks{e-mail: eugeny.babichev@th-u-psud.fr}
\affiliation{Laboratoire de Physique Theorique (UMR8627), CNRS, Univ. Paris-Sud, Universite Paris-Saclay, Orsay, 91405 France}
\author{V. I. Dokuchaev}\thanks{e-mail: dokuchaev@inr.ac.ru}
\affiliation{Institute for Nuclear Research, Russian Academy of Sciences, pr. 60-letiya Oktyabrya 7a, Moscow, 117312 Russia}
\affiliation{National Research Nuclear University MEPhI, Kashirskoe sh. 31, Moscow, 115409 Russia}
\author{A. S. Malgin}\thanks{e-mail: malgin@lngs.infn.it}
\affiliation{Institute for Nuclear Research, Russian Academy of Sciences, pr. 60-letiya Oktyabrya 7a, Moscow, 117312 Russia}

\date{\today}

\begin{abstract}
A change in gravity law in some regimes is predicted in the modified gravity models that are 
actively discussed at present. In this paper, we consider a possibility that the signal recorded by the Geograv
resonant gravitational-wave detector in 1987 during the explosion of SN 1987A was produced by an abrupt
change in the metric during the passage of a strong neutrino flux through the detector. Such an impact on
the detector is possible, in particular, in extended scalar-tensor theories in which the local matter density gradient
affects the gravitational force. The first short neutrino pulse emitted at the initial stage of stellar core
collapse before the onset of neutrino opacity could exert a major influence on the detector by
exiting the detector response at the main resonance frequency. In contrast, the influence of the subsequent
broad pulse (with a duration of several seconds) in the resonant detector is exponentially suppressed, despite
the fact that the second pulse carries an order-of-magnitude more neutrino energy, and it could generate a
signal in the LSD neutrino detector. This explains the time delay of 1.4~s between the Geograv and LSD signals.
The consequences of this effect of modified gravity for LIGO/Virgo observations are discussed.
\end{abstract}

\maketitle

\section{INTRODUCTION}

On February 23, 1987, the explosion of a core-collapse
supernova (SN) was observed in the Large Magellanic
Cloud at a distance $r\simeq52$~kpc from the Earth
(for a review and detailed discussion, see \cite{Bah93}). A statistically
significant neutrino signal from the explosion
was recorded in the Mont Blanc LSD detector (Italy,
USSR) \cite{Dadetal87}, and a second neutrino signal was recorded
by the Kamiokande II (Japan), IMB (USA), and Baksan
Neutrino Observatory (USSR) detectors 4.7~h
later. Such a double signal on a time scale of several
hours is puzzling, since the neutrino emission
lasts only a few seconds in a single collapse. 
A two-stage collapse may in principle explain this puzzle~\cite{SteTre87,Hiletal87,Ruj87,Beretal88,DadZatRya89,ImsNad88,ImsRya04,BisMoiArd18}.
Within the improved rotational mechanism described
in \cite{ImsRya04}, instability develops and two neutron stars revolving
in an orbit are formed due to the presence of a large
angular momentum in the core of the original star as it
collapses. Losing the energy of orbital motion through
gravitational radiation, the binary components
approach each other for 4.7~h. Matter from the less
massive neutron star is transferred through the Roche
point to the more massive neutron star. As a result, the
smaller (in mass) neutron star explodes, while the
larger one collapses into a black hole with the emission
of a second neutrino signal. The presence or absence
of recording of neutrino signals from different stages of
collapse by different groups of detectors can be
explained by a difference between the energy spectra
of the neutrinos generated at different evolutionary
stages of the collapsar and the detector characteristics
\cite{Beretal88} (for a review of various possibilities, see also
\cite{Bah93}, \cite{Vis15}).

Besides neutrinos, gravitational waves can also be
emitted during SN explosions. Within general relativity,
non-spherical explosion is a necessary condition
for the emission of a gravitational wave. A possibility
of non-sphericity during an explosion leading to the emission
of gravitational waves was pointed out by
L.M.~Ozernoi in 1964. 
The change in the
quadrupole moment can occur if the collapsing core is
pear-shaped. In this paper we will discuss a possibility
of recording a gravitational signal from SN~1987A
while considering the gravitational perturbation in
terms of the modified gravity theory rather than general relativity. First, let us briefly outline the history of
the question.

Several gravitational-wave signals from the mergers
of black holes and neutron stars in binary systems have
been reliably recorded with the LIGO/Virgo laser
interferometers since 2016 \cite{LIGOVirgo17-1}. However, the attempts
to record gravitational waves were made previously.
After the theoretical works by H.~Bondi and J.~Weber,
who developed the method of recording gravitational
waves by solid-state detectors, Weber constructed
such detectors in the shape of cylinders whose
oscillations were picked up with piezoelectric sensors.
A solid-state detector can record a gravitational wave if
the latter contains Fourier components close to the 
resonance frequencies of the cylinder. In 1969
Weber reported the recording of signals that could be
gravitational waves, but this result was not confirmed
in independent experiments. Nevertheless, the designs
of solid-state detectors continued to improve, while
their sensitivity increased.

Two solid-state gravitational-wave detectors operated
on February 23, 1987. A distinct signal was
recorded by the Geograv detector in Rome \cite{Amaetal87}, \cite{Agletal91}.
The probability of an accidental coincidence was estimated
in \cite{Amaetal87} to be 3\% and in a more detailed analysis
\cite{Agletal91} to be 0.001\%. This signal preceded the cluster of
neutrino pulses in the LSD detector by $1.4\pm0.5$~s. The
recording of such a signal, unless it was a statistical fluctuation,
looks puzzling, because a classical gravitational
wave could produce this signal only if an energy
of $\sim2.4\times10^3M_\odot$ was released into gravitational waves
\cite{Amaetal87}, while the entire mass of the collapsed and
exploded star, a blue supergiant, was
$\sim16M_\odot$.

Therefore it is impossible to explain the signal in the
Geograv detector in terms of general relativity. 
It is thus interesting to consider a possibility of explaining
the signal in terms of modifications of general relativity.
Some steps have
already been taken in this direction. 
For example, the effect of scalar gravitational waves on the Geograv
detector in the field theory of gravitation was investigated
in \cite{Bar97}.

In this paper we will consider the effect of a neutrino envelope passing through the Earth on
the Geograv detector. 
We carry out our study in the framework of modified gravity, in particular, as a test model
we assume beyond Horndeski theory \cite{Crisostomi:2016czh,Achour:2016rkg}. In this theory the gravitational potential contains an additional
term proportional to the local matter density
gradient\footnote{Note that such theories satisfy the local gravitational observations (for example, in the Solar system) thanks to the Vainshtein
mechanism \cite{Vainshtein} (for a review, see \cite{Babichev:2013usa}).}.  

Thus, the Geograv detector could be affected by
the density gradient of the neutrino envelope passing
through it. As we will discuss in more detail below, the
signal in the Geograv detector could be produced only
by the short (with a duration of less than $<10^{-2}$~s) neutrino
burst emitted at the initial stage of collapse, while
the main signal in the neutrino detectors should occur
about $\sim1$~s later, which gives a natural explanation for
the observed earlier signal in Geograv.

On the other hand, if the signal in the Geograv
detector on February 23, 1987, was a random fluctuation,
then it can serve to obtain upper bounds on the
parameters of the gravity theory. Such a bound is possible
even in the absence of a signal at the detector
noise level from the condition that the signal does not
exceed the noise level.

\section{THE SOURCE OF THE SIGNAL}

During the gravitational collapse of a stellar core a
strong neutrino flux comes at two stages:
at the stage of initial stellar core collapse before the
onset of neutrino opacity, when bulk neutrino emission
occurs, and at the later neutrino-opaque stage
with surface neutrino emission (for core-collapse
supernovae, see  \cite{ZasPos06}, \cite{Bah93}). Initially, the stellar core collapses
to a density $\rho_s\simeq2.8\times10^{14}$~g~cm$^{-3}$ during the time $\Delta t\sim0.001-0.01$~s with energy release $M_\nu\sim0.01M_\odot$
in the form of neutrinos. The second stage lasts much
longer, of the order of several seconds, with the release
of energy $M_\nu\sim0.1M_\odot$ in the form of neutrinos. The
neutrinosphere inevitably becomes non-spherical due to
the presence of a magnetic field and presupernova
core rotation. As a result, the duration of the initial and subsequent
stages of neutrino emission in different directions is different. 
Therefore, when the signal from the
SN is recorded, there is an additional factor which depends on the angle and that we, however, disregard here.

As we show below, the typical time scale of the neutrino pulse plays a crucial role in the possibility of gravitational signal detection.
For this reason, the second stage was unobservable for the Geograv detector. Conversely, a low neutrino flux at
the first stage makes it unobservable for neutrino telescopes.
Thus, Geograv and LSD could record the
neutrino signal from the first and second stages,
respectively. This gives a natural explanation for the
fact that Geograv recorded the signal 1.4~s earlier than
did the neutrino detector.

\begin{figure}
	\begin{center}
\includegraphics[angle=0,width=0.49\textwidth]{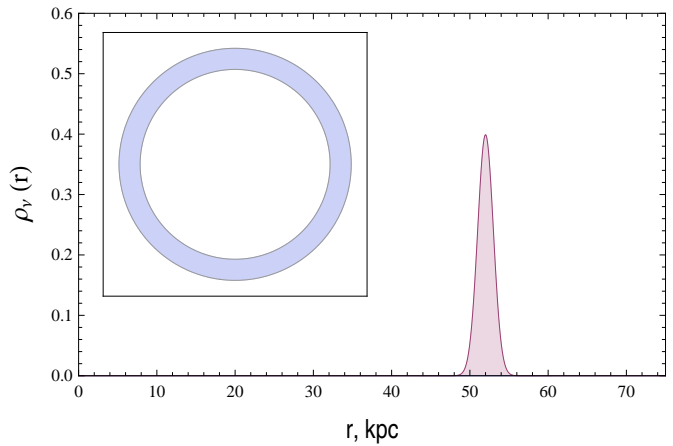}
	\end{center}
\caption{Schematic view of the neutrino envelope. The distance
$r$ is measured from the SN explosion site in the Large
Magellanic Cloud. The peak profile is given by a function
$f(r)$. The density is shown in dimensionless units; the envelope
width is increased for clarity. The metric is different inside and outside the envelope.}
	\label{gr1}
\end{figure}

Neutrinos are emitted in the form of a spherical envelope with
a characteristic thickness $H_\nu\sim c\Delta t$, see Fig.~\ref{gr1}. 
The envelope density in the laboratory frame (in
the detector frame) reads
\begin{equation}
\rho_\nu(t,r)=\frac{M_\nu}{4\pi r^2}f(r-ct),
\end{equation}
where $f(x)$ is the normalized envelope density profile at fixed $t$
(integration is along the radial direction),
\begin{equation}
\int f(x)dx=1,
\end{equation} 
and $M_\nu c^2$ is the energy of the envelope in the
laboratory frame. For simplicity we will assume that the function $f(x)$ has
the form of a Gaussian with the center at an envelope
radius $x=r(t)$ and a characteristic width $H_\nu$:
\begin{equation}
f(x)=\frac{1}{\sqrt{2\pi}H_\nu}e^{-x^2/(2H_\nu^2)}.
\label{gauss}
\end{equation}  
The real distribution can have a more complex form with different
growth and decay time scales.

Since the neutrinos have a small, but nonzero rest
mass, these particles move with a speed slightly lower
than the speed of light. As was first pointed out by
Zatsepin in \cite{Zat68}, this allows to put an upper bound
on the neutrino mass from the observations of SN explosions. 
The neutrino signal delay
compared to the speed of light is
\begin{equation}
\delta t=\frac{r}{2c}\left(\frac{m_\nu c^2}{E_\nu}\right)^2.
\label{ush1}
\end{equation}
Numerically, the quantity
\begin{equation}
\delta t=2.6\times10^{-4}\left(\frac{r}{50\mbox{~kpc}}\right)\left(\frac{E_\nu}{10~\mbox{MeV}}\right)^{-2}\left(\frac{m_\nu}{0.1~\mbox{eV}}\right)^2\mbox{~s}
\label{dtnumass}
\end{equation}
is equal in order of magnitude to the neutrino envelope
broadening if we set the typical width of the
energy spectrum equal to $\delta E_\nu\sim E_\nu$. Up-to-date data
on the neutrino masses are given in \cite{GorRub16}. The direct
experimental constraint on the electron neutrino mass
from the $\beta$-decay of tritium is $m_\nu\leq2$~eV (for Majorana
neutrinos $m_\nu\leq0.3$~eV). A constraint of the same order
of magnitude $m_\nu\leq0.2-1$~eV is obtained from the cosmological
data on the cosmic microwave background
and the formation of large-scale structures. The lower
bound on the electron neutrino mass follows from the
constraints on the difference of the squares of the
masses of mass states, which, in turn, follow from the
observations of neutrino oscillations. In the case of a
direct hierarchy of masses without degeneracy, the
electron neutrino mass is $m_\nu\geq0.009$~eV, but may also
be $m_\nu\geq0.05$~eV. At a mass $m_\nu\leq0.1$~eV the neutrino
envelope at a distance $r\sim50$~kpc will not spread out,
and its width depends only on the emission process.
This may not the case for neutrinos from other galaxies.
The spreading can become significant starting
from some distance, which  affects the detection
efficiency. The detectors may not record the neutrino
envelopes from SN explosions in distant galaxies due
to this effect.

\section{AN ADDITIONAL CONTRIBUTION TO THE GRAVITATIONAL POTENTIAL}

In this paper we make use a prediction of
beyond Horndeski theory for Geograv detector. 
The beyond Horndeski theory is a general scalar-tensor
theory containing one additional scalar
degree of freedom (in addition to graviton).
General relativity is a special case of this theory. In the
spherically symmetric case and in a quasi-static approximation
the gradient of the gravitational potential in beyond Horndeski theory reads~\cite{KobWatYam15}:
\begin{equation}
\frac{d\phi}{dr}=G\left(\frac{M(r)}{r^2}-\tilde\varepsilon\frac{d^2M(r)}{dr^2}\right),
\label{dpdrmain}
\end{equation}
where
\begin{equation}
M(r)=4\pi\int\limits_0^r r'^{2}\rho(r')dr',
\end{equation}
and $\tilde\varepsilon$ is a dimensionless parameter expressed via the fundamental
parameters of the theory and the cosmological solution\footnote{Note that in the Horndeski theory~\cite{Hor74} the second term in (\ref{dpdrmain}) is absent.}. There are, in particular, constraints on $\tilde\varepsilon\leq1$ based on the stability of neutron
stars \cite{Saietal15}. The derivative $d^2M(r)/dr^2$ has a contribution
from the density gradient. 
Note that no similar expression has been found for the case of a fast moving medium.
Therefore, strictly speaking, we cannot  apply (\ref{dpdrmain}) to the neutrino envelope.

Instead we will use a phenomenological approach
and interpret (\ref{dpdrmain}) as the result of gravity modification
that is also valid for moving matter. More specifically,
we write the additional contribution to the derivative of
the gravitational potential in the detector rest frame as
\begin{equation}
\frac{d\Delta\phi}{dr}=-\varepsilon G\frac{d^2M(r)}{dr^2},
\label{dpdrmainnew}
\end{equation}
with the parameter $\varepsilon$ being arbitrary and, generally
speaking, unrelated to any specific modified gravity
model.
As we will see below, the signal in the Geograv
detector can be explained by the presence of the
term (\ref{dpdrmainnew}). On the other hand, we will obtain an independent
constraint for the parameter $\varepsilon$.

For further purposes we will need an
estimate of $d^2M(r)/dr^2$ in (\ref{dpdrmainnew}). For the neutrino envelope
considered above we have
\begin{equation}
M(t,r)=4\pi\int\limits_0^r r^{'2}\rho_\nu(t,r')dr',
\end{equation}
\begin{equation}
\frac{\partial M(r)}{\partial r}=4\pi r^2\rho_\nu(t,r)=M_\nu f(t,r),
\end{equation}
\begin{equation}
\frac{\partial ^2M(r)}{\partial r^2}=M_\nu\frac{\partial f(t,r)}{\partial r}\sim\frac{M_\nu}{H_\nu},
\end{equation}
since the function $f(t,r)$ changes on the length
scale $H_\nu$.

\section{THE SIGNAL IN THE DETECTOR}

Consider a solid-state detector of a cylindrical shape of length $L$ and whose axis and the the plane of the incoming neutrino envelope make an angle $\theta$. 
The basics of operation of such detectors is described in
detail in \cite{AstKvaTeo82}. A gravitational wave or other action on
the cylinder causes its oscillations that are recorded
with the piezoelectric elements fixed to the detector.
We will place the origin of coordinates at the cylinder center;
the ``z'' axis is directed along the cylinder. The tidal
acceleration acting on a cylinder mass element with
coordinate $z$ reads
\begin{equation}
g_t=\frac{\partial \phi(t,r)}{\partial r}z\sin\theta,
\end{equation}
where $\theta$  was $30^\circ$ at the time of observation of the first
neutrino signal from SN~1987A. The extra contribution to the gravitational potential~(\ref{dpdrmainnew}) is given by
\begin{equation}
g_t=-z\sin\theta\varepsilon GM_\nu\frac{\partial^2 f(t,r)}{\partial r^2}.
\end{equation}
Note that the ordinary Newtonian force contains
$1/r^2$ factor, where $r=52$~kpc, and, therefore, its contribution
to the signal is negligible with respect to the noise level.

The equation for longitudinal cylinder oscillations
with a tidal force reads~\cite{LLUprug}
\begin{equation}
\rho\frac{\partial^2u_z}{\partial t^2}=E\frac{\partial^2u_z}{\partial z^2}+\left(\frac{4}{3}\eta+\zeta\right)\frac{\partial}{\partial t}\frac{\partial^2u_z}{\partial z^2}+\rho g_t,
\label{oscbigeq}
\end{equation}
where $u_z$ is a vector component, $\rho$ is the density, $E$ is
Young's modulus, $\eta$ and $\zeta$ are the viscosity coefficients
of the cylinder material defining the damping
time of its oscillations. 
Note that for $f(t,r)=f(r-ct)$ we have $\partial^2 f/\partial r^2=(1/c^2)\partial^2 f/\partial t^2$. 
Therefore, given the above relations, the term $\rho g_t$ on the right-hand-side of (\ref{oscbigeq}) can be written as
\begin{equation}
\rho g_t=\frac{\rho}{2}z\frac{\partial^2}{\partial t^2}\left[2\varepsilon\sin\theta\frac{GM_\nu}{c^2}f\right].
\label{rhogt}
\end{equation}
We see that Eq.~(4.5) from \cite{AstKvaTeo82} coincides with (\ref{oscbigeq})
after the substitution
\begin{equation}
h\to 2\varepsilon\sin\theta\frac{GM_\nu}{ c^2}f.
\label{replace}
\end{equation}
Thus, the response in the detector can be taken into
account by the same method \cite{AstKvaTeo82} that was applied in
the case of gravitational waves in general relativity,
but with the substitution of the combination of
quantities (\ref{replace}) for the wave amplitude. The detector
response is expressed via the signal spectral density
close to resonances. Since the signal frequency is lower
than the first resonance frequency of the Geograv
detector in the case under consideration, the region
near the first resonance frequency $\omega_0$ plays a major
role. 
Close to the resonance a cylindrical detector is equivalent
in response to a system of two weights connected
by a spring. When choosing (\ref{gauss}), the Fourier
transform of the quantity (\ref{replace}) is
\begin{equation}
H(\omega)=2\varepsilon\sin\theta\frac{GM_\nu}{c^3}e^{-i\omega r/c}e^{-H_\nu^2\omega^2/(2c^2)}.
\label{hbig}
\end{equation}
The cylinder oscillations resulting from the envelope
passage are found by the convolution
\begin{equation}
u_z(t,z)=\frac{1}{2\pi}\int\limits_{-\infty}^\infty T(z,\omega)H(\omega)e^{i\omega t}d\omega,
\label{intht}
\end{equation}
where $T(z,\omega)$ is the response function obtained by the
Fourier transform of Eq.~(\ref{oscbigeq}).

The response function of the equivalent spring
detector near the first resonance can be approximately
given as~\cite{AstKvaTeo82}
\begin{equation}
T(\omega)=\frac{2L}{\pi^2}\frac{\omega^2}{\omega^2-\omega_0^2-i\omega/\tau_0},
\label{tbig}
\end{equation}
where $\tau_0=Q/\omega_0$ is the oscillation damping time and $Q$
is the detector $Q$-factor. We obtain the motion
for a weight of the equivalent spring detector at $t>r/c$ by
calculating the integral (\ref{intht}):
\begin{eqnarray}
\xi(t)&=&\frac{4L\omega_0}{\pi^2}\varepsilon\sin\theta\frac{GM_\nu}{c^3}\sin\left[\omega_0(t-r/c)\right]\times \nonumber
\\
\nonumber
\\
&\times& \exp\left\{\frac{-(t-r/c)}{2\tau_0}\right\}\exp\left\{-H_\nu^2\omega_0^2/(2c^2)\right\}.
\end{eqnarray}

The energy released in the detector does not exceed
the spike in the effective temperature recorded on
February 23, 1987:
\begin{equation}
E_{\rm max}=\frac{1}{4}M\omega_0^2\xi_{\rm max}^2\leq\frac{k_BT^*}{2},
\label{emax}
\end{equation}
where $M$ is the cylinder mass (the mass of each of the
two weights of the equivalent detector is equal to $M/2$)
and $T^*\sim 135$~K is the temperature spike (at a noise
level of 29~K); the equality in (\ref{emax}) will hold if the spike
in temperature is explained by the passage of the neutrino
envelope. Hence we obtain
\begin{equation}
\varepsilon\leq\frac{\pi^2(k_BT^*)^{1/2}c^3}{2^{3/2}GM_\nu\omega_0^2M^{1/2}L\sin\theta}e^{H_\nu^2\omega_0^2/(2c^2)}.
\label{itogrestr}
\end{equation}
This constraint is shown in Fig.~2.

The envelope width $H_\nu=c\Delta t$ dependent on the
neutrino signal duration $\Delta t$ plays a crucial role in exponential factor in (\ref{itogrestr}). The duration of the
first neutrino emission stage coincides in order of
magnitude with the free-fall time $t_{\rm ff}\sim(G\rho_s)^{-1/2}\sim 4\times10^{-3}$~s at $\rho_s\sim10^{12}$~g~cm$^{-3}$. Numerical SN explosion
simulations \cite{Nad78} show a neutrino burst on a characteristic
time scale $\Delta t\sim10^{-3}$~s due to the sharp growth in
plasma temperature during the collapse. This primary
burst continues until the onset of neutrino opacity, the
optical depth for neutrinos $\tau_\nu=1$. 
It can be seen from Fig.~2 that the signal in the Geograv detector can be generated for $\Delta t\ll 10^{-2}$~s. 
The most conservative constraint
$\tilde\varepsilon\leq1$ \cite{Saietal15} based on the stability of neutron stars at $\varepsilon\sim\tilde\varepsilon$
would rule out a duration $\Delta t\geq10^{-2}$~s.

The second neutrino burst (at the first stage of two-stage
collapse) after the onset of neutrino opacity lasts
much longer, of the order of several seconds, and,
therefore, its influence on the detector is exponentially
suppressed. The neutrino luminosity in the second
burst probably decreases exponentially and not as a
Gaussian, but only the characteristic decline time is
important for a qualitative estimate. Meanwhile,
during the second burst an order of magnitude greater
energy, $6.3\times10^{53}$~erg$=0.35M_\odot c^2$ \cite{Beretal88}, is released into
neutrinos and, therefore,
the signals in the neutrino detectors are due to
to the second burst. This explains the delay of
1.4~s between the signal recordings by the Geograv
detector and the LSD neutrino detector. However, it
should be noted that the time in the neutrino detector
is measured from the first neutrino event and, therefore, this delay could be slightly different in view of the
statistical fluctuations.

During two-stage collapse one expects additional
neutrino signals 4.7~h later that were recorded by the
Kamiokande II, IMB, and Baksan detectors. 
In order to estimate the signal in the gravitational-wave detector from the second stage 
one needs to know the duration of the neutrino emission that is generated as the neutron
star fragments collide \cite{Beretal88} or as the neutron star collapses into a black hole. 
First of all, it can be seen from
the observational data (all neutrino events are listed in
\cite{Bah93}) that the detectors recorded the neutrino signal for
several seconds: up to 12.4 and 5.6~s in the case of the
Kamiokande II and IMB detectors, respectively.
Thus, the main neutrino flux constituted an extended
envelope the signal from which in the Geograv detector
is exponentially suppressed, as seen from Fig.~2.
The following question remains: could more rapid
variations be present in the overall extended signal, as
is the case during an ordinary core-collapse SN explosion,
when the first short neutrino burst is generated
during the primary stellar core collapse? If the second
stage is associated with the merger of two neutron stars
or a black hole and a neutron star, then the data on
short cosmic gamma-ray bursts, which are now known
to be generated in such mergers, may turn out to be
useful (though no gamma-ray burst was observed
during the explosion of SN~1987A). The variability of
some short gamma-ray bursts occurs on a time scale of
$\sim10^{-3}$~s. If the neutrino envelopes could have the same
time scale, then they could be recorded by the Geograv
detector. In this case, the question about the
absence of recording of a second gravitational signal
remains unresolved. At the same time, other calculations
\cite{Becetal10} suggest that the typical variability time in the
fireball of a gamma-ray burst $\Delta t$ is closer to $\sim10^{-2}$~s,
while many short bursts last up to 3 s, and the signal in
the detector will then be exponentially suppressed,
Thus, the absence of a second Geograv signal concurrent
with Kamiokande II and IMB can be explained
either by the absence of rapid variability, i.e., a large
$\Delta t\geq10^{-2}$~s, or by minor energy release into neutrinos
in episodes of rapid variability.

\begin{figure}
	\begin{center}
\includegraphics[angle=0,width=0.49\textwidth]{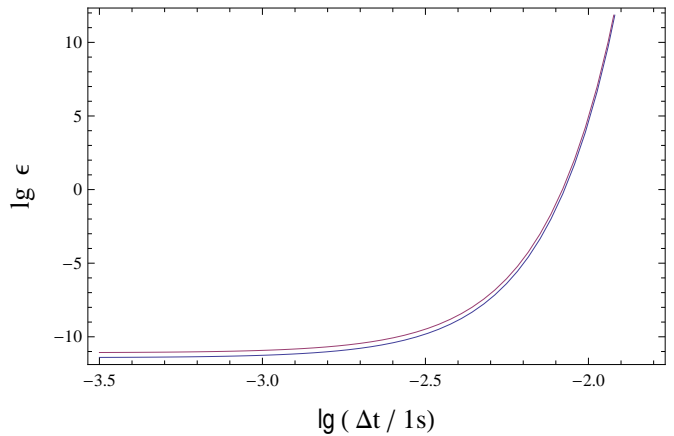}
	\end{center}
\caption{Constraint on the parameter $\varepsilon$ versus initial neutrino
burst duration $\Delta t$ at $M_\nu=0.02M_\odot$. The lower and upper curves correspond to a
noise level $T^*\simeq29$~K and a possible signal level $T^*\simeq135$~K,
respectively.}
	\label{gr2}
\end{figure}

Note that the distance to the source does not enter
explicitly into the condition (\ref{itogrestr}). Therefore, one
would think that the detectors should record the signals
from SN explosions in other galaxies very frequently.
However, (\ref{itogrestr}) can depend on the distance via
$\Delta t$, which increases with distance due to a nonzero
neutrino mass, according to Eq.~(\ref{dtnumass}). This effect can
explain the absence of signals from distant galaxies. At
even greater distances, $\sim1$~Gpc, the influence of a cosmological
redshift should manifest itself.

\section{ESTIMATING THE SIGNAL IN A GRAVITATIONAL INTERFEROMETER}

Above we calculated the signal in a resonant solid-state
detector. Let us now estimate the signal that
could be recorded by gravitational interferometer with
free mirrors if a SN exploded during
their operation. This question can be of importance
for LIGO/Virgo observations and future gravitational detectors, since sooner or later another corecollapse
SN explosion will occur in our Galaxy or its
surroundings. According to various estimates, such
explosions occur once in 20100 years. Let us consider
an interferometric detector in the form of two
free masses spaced a distance $L$ apart. The neutrino
envelope crosses the detector during a characteristic time
$\Delta t\sim H_\nu/c$. A tidal acceleration
\begin{equation}
a\sim\frac{d^2\phi}{dr^2}L\sim G\varepsilon L\frac{M_\nu}{H_\nu^3}
\label{atid}
\end{equation} 
acts during this time. Since the envelope moves relativistically,
a contribution of not only the energy, but
also the momentum of the envelope will be present in
the gravitational force. However, this contribution is
of the same order of magnitude as (\ref{atid}) and we disregard
it in our order-of-magnitude estimate. The
change in interferometer arm in the envelope passage
time is estimated as $\delta L\sim a(\Delta t)^2/2$ and the relative
change in the size is
\begin{equation}
h=\frac{\delta L}{L}\sim\frac{GM_\nu\varepsilon}{2H_\nu c^2}\sim \frac{\varepsilon}{4}\frac{r_g}{H_\nu},
\end{equation} 
where the gravitational radius $r_g=2GM_\nu/c^2$. Numerically,
we have
\begin{equation}
h\simeq 5\times10^{-6}\left(\frac{r_g}{0.02\times3\mbox{~km}}\right) \left(\frac{\Delta t}{10^{-2}\mbox{~s}}\right)^{-1}\varepsilon
\label{hrez}
\end{equation} 
This quantity at $\varepsilon>10^{-15}$ can exceed the effect of the gravitational wave.

If there is a broadening of the neutrino envelope
due to a nonzero neutrino mass, according to Eq.~(\ref{ush1}),
then in the above calculation we should substitute the
following quantity as $c\Delta t$:
\begin{equation}
\Delta t={\rm max}\left\{H_\nu,\quad \frac{r}{2}\left(\frac{m_\nu c^2}{E_\nu}\right)^2\right\}.
\end{equation}
In our calculation we also assumed the envelope width
to be $c\Delta t\geq L$. For future space interferometers like
LISA this condition may not be satisfied. In the case
of $c\Delta t\ll L$, the tidal acceleration (\ref{atid}) will be gained on
a scale of the order of $\sim c\Delta t$ and, therefore, the additional
small factor $c\Delta t/L$ will enter into (\ref{hrez}).

Each interferometric detector has its sensitivity
curve; it can receive signals only in a limited frequency
range. Therefore, for an excessively long pulse (low
characteristic frequency) the signal will be outside the
detection range. The characteristic signal frequency
\begin{equation}
\nu\sim\frac{c}{H_\nu}\sim\frac{1}{\Delta t}=10^2 \left(\frac{\Delta t}{10^{-2}\mbox{~s}}\right)^{-1}\mbox{~Hz}
\end{equation}
at $\Delta t$ specified in the normalization falls into the most
sensitive range of the LIGO/Virgo detectors. Therefore,
during the explosion of a supernova like SN~1987A it would be possible to obtain a strong
constraint on $\varepsilon$.

In 2017 the LIGO/Virgo detectors recorded the
gravitational-wave burst GW170817 [12]. The short
gamma-ray burst GRB 170817A was recorded from its
localization region by the Fermi-GBM telescope
$1.74\pm0.05$~s later. This event is most likely associated
with the merger of two neutron stars or a neutron star
and a black hole with masses of $1.17-1.60M_\odot$. A
powerful neutrino signal must be generated during
such a merger. According to optical observations, the
source is located in the galaxy NGC~4993 at a distance
of $40\pm8$~Mpc from the Earth \cite{LIGOVirgo17-2}, i.e., farther than
SN~1987A approximately by a factor of $\sim10^3$. The
absence of anomalies in the LIGO/Virgo observations
can be explained by one of the following factors or
their combination: a small parameter $\varepsilon\ll10^{-15}$, a large
$\Delta t\gg10^{-2}$~s or the influence of the neutrino mass and
the spreading of the neutrino envelope to large $\Delta t$ as
the neutrinos traverse a distance of $40\pm8$~Mpc
(according to Eq.~(\ref{dtnumass})) at $m_\nu\geq0.01$~eV.

Note also that the Newtonian part of the tidal
acceleration produced by the passing envelope is
\begin{equation}
a_{\rm N}\sim 2\pi G\frac{M_\nu}{4\pi r^2}\frac{L}{H_\nu}
\end{equation}
and produces a single pulse with an amplitude
\begin{eqnarray}
h&=&\frac{\delta L}{L}\simeq 10^{-56}\left(\frac{r_g}{0.02\times3\mbox{~km}}\right)\times \nonumber
\\
\nonumber
\\
&\times& \left(\frac{\Delta t}{10^{-2}\mbox{~s}}\right)\left(\frac{r}{52\mbox{~kpc}}\right)^{-2},
\end{eqnarray}
whose value is far too small to be detected.

\section{CONCLUSIONS}
In this paper we showed that the signal probably recorded in the
Geograv gravitational-wave detector during the explosion
of SN~1987A \cite{Amaetal87} could have been produced by a gravitational
field perturbation during the passage of a powerful
neutrino flux through the detector.
The appearance of such a strong signal
is possible in modern scalar-tensor gravity theories,
like extensions of Horndeski theory, in which
the gravitational potential depends not only on the
objects mass and distance, but also on the local matter
density gradient. This model successfully explains
the time delay of 1.4~s between the Geograv and LSD
signals. The signal in Geograv is attributable to the
neutrinos emitted at the initial stage before the onset
of neutrino opacity, while the signal in LSD was produced
by the main neutrino flux.

The signal from the neutrino pulse has different
polarization compared to a
gravitational wave in general relativity, and it produces oscillations in
a solid-state resonant detector. Yet
another difference  
is that the additional effect due to the neutrino pulse is not of the form of
oscillations, contrary to the case when gravitational waves in general relativity or scalar gravitational
waves~\cite{Bar97} are emitted by an oscillating body.

We also showed that LIGO/Virgo gravitational interferometers would see 
such signals when the neutrino pulses exploded supernovae were passing through detectors.

It should be stressed that exact solutions for the gravitational potentials for a fast moving medium in beyond Horndeski theory have not been found yet. In this paper we use a phenomenological
approach by introducing an unknown parameter $\varepsilon$ that
relates the potential gradient to the envelope density
gradient in the detector rest frame, by extending 
the static solution~\cite{KobWatYam15,Saietal15} to the case of moving neutrinos. 
With this assumptions we demonstrated a possibility to
explain the signal in the Geograv detector and we found exponential dependence of the
effect on the neutrino signal duration.

It would be of interest to also consider the collapse in modified gravity when simulating the
gravitational pre-supernova core collapse, and to calculate the shape of the neutrino signal.
This would allow to study self-consistently the problem of the
detection of gravitational signals in such theories.

The question of whether an additional contribution in gravitational potential 
is present when a pulse of gravitational waves passes is also of interest\footnote{We are grateful to the referee of this paper who pointed to this possibility.}. Such an pulse does not
spread out and, therefore, the signal can arrive from
great distances. In beyond Horndeski theory 
the matter energy-momentum tensor makes an additional
contribution to the potential gradient. 
It would be interesting to see whether a gravitational wave, i.e., the energy-momentum pseudo-tensor of the gravitational field,
can make a similar contribution.

{\bf Acknowledgments}

This research was supported by the
CNRS/RFBR Cooperation program for 2018-2020
n.~1985 (France), Russian Foundation for Basic
Research (project n.~18-52-15001 CNRS) (Russia)
Modified gravity and black holes: consistent models and experimental signatures, and
the research program ``Programme national de cosmologie et galaxies'' of the CNRS/INSU.


\begin{thebibliography}{99}

\bibitem{Bah93} J. Bahcall, {\it Neutrino Astrophysics} (Cambridge Univ.
Press, Cambridge, 1989).

\bibitem{Dadetal87} V.~L.~Dadykin, G.~T.~Zatsepin, V.~B.~Korchagin et al., JETP Lett. {\bf 45}, 593 (1987).

\bibitem{SteTre87} L.~Stella and L.~Treves, Astron. Astrophys. {\bf 185}, L5 (1987).

\bibitem{Hiletal87} W.~Hillebrandt, P.~Hoflich, P.~Kafka, Astronomy and Astrophysics, {\bf 180}, L20 (1987).

\bibitem{Ruj87} A.~De~Rujula, Phys. Lett. B {\bf 193}, 514 (1987).

\bibitem{Beretal88} V.~S.~Berezinsky, C.~Castagnoli, V.~I.~Dokuchaev et al., Il Nuovo Cimento {\bf 11}, 287 (1988).

\bibitem{DadZatRya89} V. L. Dadykin, G. T. Zatsepin, and O. G. Ryazhskaya, Sov. Phys. Usp. {\bf 32}, 139459 (1989).

\bibitem{ImsNad88}  V. S. Imshennik and D. K. Nadezhin, Usp. Fiz. Nauk {\bf 156}, 561 (1988).

\bibitem{ImsRya04} V. S. Imshennik and O. G. Ryazhskaya, Astron. Lett. {\bf 30}, 14 (2004).

\bibitem{BisMoiArd18} G. S. Bisnovatyi-Kogan, S. G. Moiseenko, and
N. V. Ardelyan, Phys. At. Nucl., {\bf 81}, 266 (2018).

\bibitem{Vis15} F.~Vissani, Journal of Physics G: Nuclear and Particle Physics {\bf 42}, 013001 (2015); arXiv:1409.4710 [astro-ph.HE].

\bibitem{LIGOVirgo17-1} B.~P.~Abbott, R.~Abbott, T.~D.~Abbott et al., Phys. Rev. Lett. {\bf 119}, 161101 (2017); arXiv:1710.05832 [gr-qc].

\bibitem{Amaetal87} E.~Amaldi P.~Bonifazi, M.~G.~Castellano et al., Europhysics Letters {\bf 3}, 1325 (1987).

\bibitem{Agletal91} M.~Aglietta, A.~Castellina, W.~Fulgione et al., Nuovo Cimento C {\bf 14}, 171 (1991).

\bibitem{Bar97} Yu.~V.~Baryshev, Astrophysics {\bf 40}, 244 (1997).

\bibitem{Crisostomi:2016czh} M.~Crisostomi, K.~Koyama and G.~Tasinato, JCAP {\bf 1604}, 044 (2016); arXiv:1602.03119 [hep-th].

\bibitem{Achour:2016rkg} J.~Ben Achour, D.~Langlois and K.~Noui, Phys. Rev. D {\bf 93}, 124005 (2016); arXiv:1602.08398 [gr-qc].

\bibitem{Vainshtein} A. I. Vainshtein, {\it Phys. Lett.} B {\bf 39}, 393 (1972).

\bibitem{Babichev:2013usa}  E.~Babichev and C.~Deffayet, Class. Quant. Grav.  {\bf 30}, 184001 (2013); arXiv:1304.7240 [gr-qc].
 
\bibitem{KobWatYam15} T.~Kobayashi, Y.~Watanabe, D.~Yamauchi, Phys. Rev. D {\bf 91}, 064013 (2015); arXiv:1411.4130 [gr-qc].

\bibitem{Saietal15} R.~Saito D.~Yamauchi, S.~Mizuno et al., Astroparticle Physics {\bf 06}, 008 (2015); arXiv:1503.01448 [gr-qc].

\bibitem{ZasPos06} A. V. Zasov and K. A. Postnov, {\it General Astrophysics}, (Vek 2, Fryazino, 2016) [in Russian].

\bibitem{Zat68} G. T. Zatsepin, JETP Lett. {\bf 8}, 205 (1968).

\bibitem{GorRub16} D. S. Gorbunov and V. A. Rubakov, {\it Introduction to the Theory of the Early Universe: Hot Big Bang Theory}, (World Scientific, 2011).

\bibitem{Hor74} G.W.~Horndeski, Int. J. of Theor. Phys. {\bf 10}, 363 (1974). 

\bibitem{AstKvaTeo82} Amaldi and G. Pizzella in {\it Astrophysics, Quanta and
Theory of Relativity}, Collection of Articles (Mir, Moscow, 1982), p. 241 [in Russian].

\bibitem{LLUprug} L. D. Landau and E. M. Lifshitz, {\it Theory of Elasticity} (Pergamon Press, New York, 1986).

\bibitem{Nad78} D.~K.~Nadezhin, Astrophysics and Space Science {\bf 53}, 131 (1978).

\bibitem{Becetal10} J.~K.~Becker, F.~Halzen, A.~O'Murchadha et al., arXiv:1003.4710 [astro-ph.HE].

\bibitem{LIGOVirgo17-2} B.~P.~Abbott, R.~Abbott, T.~D.~Abbott et al., Astrophys. J. Lett. {\bf 848}, L12 (2017); arXiv:1710.05833 [astro-ph.HE].

\end{thebibliography}
\end{document}